\documentclass[aps,prl,twocolumn,superscriptaddress,showpacs,preprintnumbers,amsmath,amssymb]{revtex4}

\usepackage{graphicx}
\usepackage{bm}

\begin{document}

\title{Two-Dimensional Magnetic Correlations and Partial Long-Range Order in Geometrically Frustrated Sr$_{2}$YRuO$_{6}$}

\author{E. Granado}
\email{egranado@ifi.unicamp.br}
\affiliation{Instituto de F\'{i}sica ``Gleb Wataghin,'' Universidade Estadual de Campinas, Campinas, S\~{a}o Paulo 13083-859, Brazil}

\author{J. W. Lynn}
\affiliation{NIST Center for Neutron Research, National Institute of Standards and Technology, Gaithersburg, Maryland 20899, USA}

\author{R. F. Jardim}
\affiliation{Instituto de F\'{i}sica, Universidade de S\~{a}o Paulo, CP 66318, S\~{a}o Paulo 05315-970, Brazil}

\author{M. S. Torikachvili}
\affiliation{Department of Physics, San Diego State University, San Diego, California 92182, USA}

\begin{abstract}

Neutron diffraction on the double perovskite Sr$_{2}$YRuO$_{6}$ with a quasi-face-centered cubic (FCC) lattice of Ru moments reveals planar magnetic correlations that condense into a partial long-range ordered state with coupled alternate antiferromagnetic (AFM) YRuO$_{4}$ square layers coexisting with short-range correlations below $T_{N1}=32$ K. A second transition to a fully ordered AFM state below $T_{N2}=24$ K is observed. These observations are ascribed to the cancellation of the magnetic coupling between consecutive AFM square layers in FCC antiferromagnets, notably the simplest three-dimensional frustrated magnet model system. Inhomogeneous superconductivity in Sr$_{2}$YRu$_{1-x}$Cu$_x$O$_{6}$, reported in earlier works, may be nucleated in magnetically disordered YRu$_{1-x}$Cu$_x$O$_{4}$ layers with strong spin fluctuations.

\end{abstract}


\maketitle

Geometrically frustrated magnets are fascinating materials displaying a rich variety of physical states. In these systems, magnetic moments interact by exchange, but no long-range magnetic structure is able to satisfy all antiferromagnetic (AFM) interactions simultaneously \cite{Lacroix}. Long-range magnetic order, if present at all, occurs only for $T < T_N << |\Theta_{CW}|$, where $\Theta_{CW}$ is the Curie-Weiss temperature, while exotic correlated magnetic states may occur for $T_N < T < |\Theta_{CW}|$. The simplest three-dimensional structure leading to frustrated magnetism and the first one to be investigated is the face-centered cubic (FCC) lattice with antiferromagnetic (AFM) nearest-neighbor (n.n.) interactions \cite{Li}. In this case the magnetic order is theoretically predicted to be unstable down to the lowest temperatures as long as both next-nearest-neighbor (n.n.n.) interactions and magnetic anisotropy are negligible \cite{Li,Kuzmin}. Despite the fundamental interest in the FCC antiferromagnets as one of the primordial examples of frustrated magnetism, the short-range correlations in this lattice have not been investigated in detail, to the best of our knowledge, leaving a gap in the mapping and understanding of the behavior of frustrated magnets.

\begin{figure}
\includegraphics[width=0.8\textwidth]{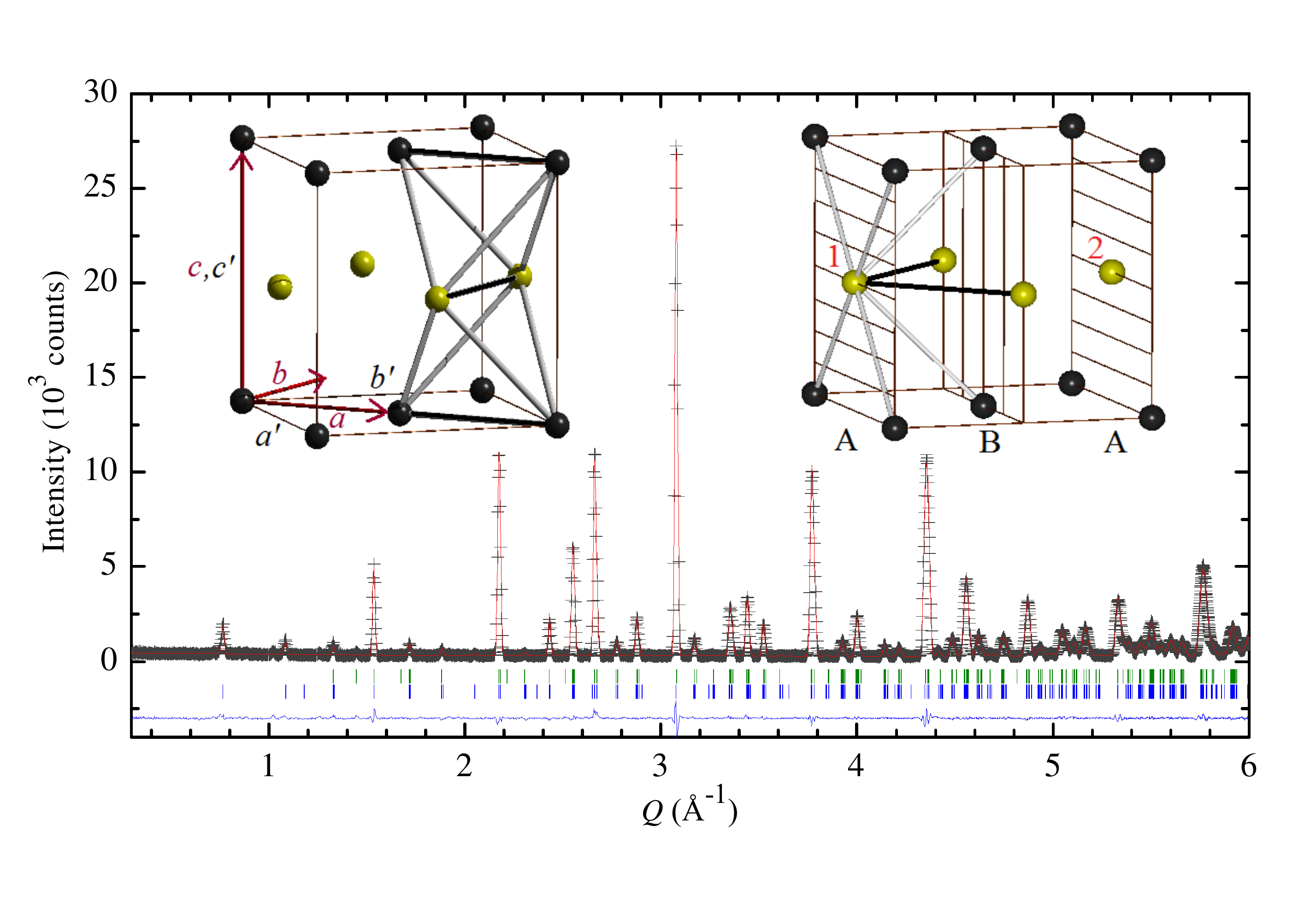}
\vspace{-1.0cm}
\caption{\label{profiles} High resolution neutron powder diffraction pattern of Sr$_2$YRuO$_6$ at 3 K. The cross symbols and solid lines represent observed and calculated patterns, respectively. The difference curve is shown at the bottom. Vertical bars indicate the expected Bragg peak positions according to the nuclear (upper) and magnetic (lower) structure models described in the text and refined parameters given in Ref. \cite{SM}. The insets illustrate the observed type-I antiferromagnetic (AFM) structure of Sr$_2$YRuO$_6$, where only the magnetic Ru ions are displayed and different shades (colors) represent opposite magnetic moments. Two representative edge-shared Ru tetahedra, which are the basic units leading to geometric frustration in this system, are illustrated in the left, also showing the monoclinic axes with lengths {\it a, b,} and {\it c} and quasi-cubic unit cell ($a' \sim b' \sim c'$). Satisfied (frustrated) nearest-neighbor AFM interactions are represented as gray (black) bonds. In the right, the same magnetic structure is visualized in terms of consecutive AFM square layers in the {\it ABAB} sequence (shaded). The net magnetic coupling between consecutive {\it AB} layers is null, while alternate {\it AA} layers are coupled by weak next-nearest-neighbor exchange interactions such as those between atoms 1 and 2 (see text).} 
\vspace{0.5cm}
\end{figure}

Sr$_2$YRuO$_6$ (SYRO) crystallizes in the ordered double perovskite structure with Ru$^{5+}$ ions defining an FCC magnetic network, which orders in a type-I AFM structure at low-$T$ \cite{Battle} (see Fig. \ref{profiles}). The magnetic order may be stabilized by weak anisotropy and/or n.n.n. interactions, which are three orders of magnitude weaker than the frustrated n.n. interactions \cite{Kuzmin}. Specific heat and magnetic susceptibility measurements indicate two phase transitions, at  $\sim 30$ K and $\sim 26$ K \cite{Singh}, while the physical state of SYRO between these temperatures still remains to be elucidated. Additional interest in this material has been generated by the occurrence of superconductivity when Ru is partially ($< 15$ \%) replaced by Cu in both powders \cite{Wu1,Wu2,Harshman1,DeMarco,Blackstead1,Blackstead2,Harshman2} and single crystals \cite{Rao1,Rao}, with superconducting volume fractions so far below 10 \%. The scenarios invoked to explain superconductivity in this system are (i) a superconducting hole-condensate in the nonmagnetic SrO layers \cite{Blackstead2} and (ii) the possibility of an impurity phase of YSr$_2$Cu$_3$O$_7$ \cite{Galstyan,Galstyan2} - which was not confirmed at least for some superconducting samples \cite{Blackstead3}. On the other hand, the superconducting temperature $T_c$ seems to be close to the magnetic ordering temperature \cite{Harshman2,Rao1,Rao}, suggesting that magnetism might play a role in the superconducting pairing mechanism. In this work, we performed a detailed investigation of the crystal and magnetic structure and correlations of the parent compound SYRO by means of high resolution and high intensity neutron powder diffraction. Our results indicate that the short-range magnetic correlations above $T_{N2}=24$ K are two-dimensional (2D) and the phase between $T_{N2}$ and $T_{N1}=32$ K is a partially ordered state with alternated AFM ordered and disordered square layers. The 2D nature of the correlations and the possibility of a partially ordered phase are argued to be general trends in FCC frustrated antiferromagnets. A scenario for superconductivity in Sr$_{2}$YRu$_{1-x}$Cu$_x$O$_{6}$ is proposed, where an inhomogeneous superconductor state is formed within magnetically disordered YRu$_{1-x}$Cu$_x$O$_{4}$ layers with strong short-range correlations.

The 13 g SYRO ceramic sample employed in this work was synthesized by solid-state reaction. Stoichiometric amounts of high-purity Ru, Y$_2$O$_3$ and SrCO$_3$ were ground together thoroughly in an agate mortar, placed in an alumina crucible, and fired first at 900 $^{\circ}$C for three days in air, with two intermediate regrindings. The resulting powder was ground again, pressed into pellets and heat treated in air at 1350 $^{\circ}$C for three days. High resolution neutron powder diffraction experiments were carried out in the BT-1 powder diffractometer of NIST Center for Neutron Research (NCNR), using a Ge(311) monochromator with $\lambda=2.0783$ \AA\ and collimation of 60' before monochromator, 20' before sample and 7' before detectors. High intensity neutron scattering measurements were performed in the BT-7 double focusing triple-axis spectrometer of NCNR operated in two-axis mode using a pyrolitic graphite monochromator with $\lambda = 2.35$ \AA\ and a position-sensitive detector \cite{BT7}. In both setups, the pellets were placed inside a vanadium can attached to a high cooling power closed-cycle He cryostat. Rietveld refinements were performed using the program GSAS+EXPGUI \cite{GSAS} under the monoclinic space group $P2_1/n$ with initial parameters taken from Ref. \cite{Battle}. Weak peaks due to a minor Y$_2$O$_3$ impurity phase (0.14 \% weight fraction) and other unindexed reflections had intensities below 0.5 \% of the main reflections, indicating a high purity level of our sample. In this paper, an $(hkl)$ notation is employed for reflection indexing according to the primitive monoclinic unit cell, while $(hkl)'$ indexing refers to the quasi-cubic unit cell (see Fig. \ref{profiles}).

Figure \ref{profiles} shows the high-resolution neutron powder diffraction pattern of SYRO at 3 K as a function of $Q \equiv (4 \pi / \lambda) \sin \theta$. Besides the nuclear Bragg peaks, magnetic reflections were observed below $T_{N1}=32$ K. These contributions were initially modeled as a type-I AFM long-range ordering (LRO) of the Ru magnetic moments, in accordance with ref. \cite{Battle}. A relevant improvement in the fitting was obtained by including oxygen moments in the magnetic structure model, as suggested by Mazin and Singh based on first principle calculations \cite{Mazin}. The oxygen polarization was found to have the same sign of the nearest Ru moment, consistent with a Ru $4d$ - O $2p$ hybridization mechanism. In addition, our resolution was sufficient to resolve the positions of the (100) and (010) neighboring reflections, revealing contributions solely from (100). The absence of the (010) magnetic reflection allows us to unequivocally determine the moment direction to be along the principal monoclinic axis [010] for all studied temperatures in the ordered phase. Indeed, the intensity ratio between the main magnetic reflections remains unaltered below $T_{N1}$. The calculated pattern at 3 K after crystal and magnetic structure refinements is given in Fig. \ref{profiles}, showing very good agreement with the observed data. The refined magnetic moment magnitudes are 1.96(2) $\mu_B$/Ru and 0.056(6) $\mu_B$/O at 3 K, which are comparable to 1.7 $\mu_B$/Ru and 0.10 $\mu_B$/O predicted by first principle calculations \cite{Mazin}. Error bars in parentheses are statistical only and represent one standard deviation. The high sensitivity of our experiment to O moments arises from interference of the scattering by the six O moments per formula unit with the contribution by the Ru moments, leading to measurable effects in the relative Bragg intensities. The similar Ru and Y scattering lengths for both neutrons and x-rays prevent a direct estimation of the Ru/Y anti-site disorder \cite{Aharen}. Nonetheless, the very different refined Ru-O and Y-O average bond distances ($1.954$ and 2.206 \AA, respectively, see \cite{SM}) and the reasonably small oxygen Debye-Waller parameters ($U_{iso} \sim 0.005$ \AA$^2$) at low-$T$ indicate a negligible degree of Y/Ru chemical disorder, which is favoured by the largely different Shannon radii of Y$^{3+}$ and Ru$^{5+}$ ions \cite{Shannon,Woodward,Barnes}. No structural phase transition is observed between 3 and 45 K, although small lattice parameter anomalies are noticed below $T_{N1}$ \cite{SM}.

\begin{figure}
\includegraphics[width=0.7\textwidth]{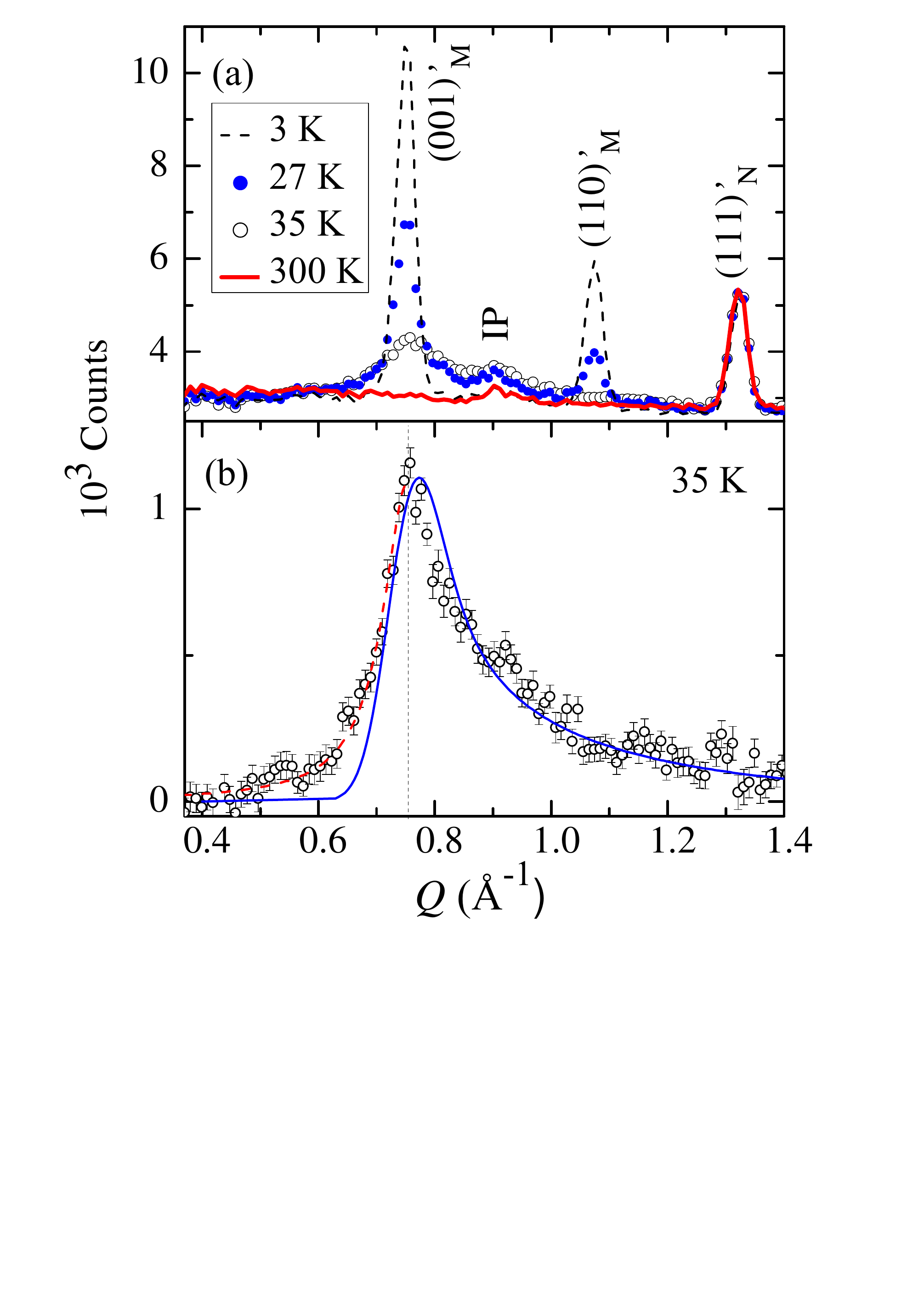}
\caption{\label{lowQ} (a) High-intensity diffraction pattern of Sr$_2$YRuO$_6$ at selected temperatures. The indexing of magnetic and nuclear Bragg reflections according to the quasi-cubic unit cell of Fig. \ref{profiles} is indicated, as well as a weak peak due to a minor unidentified impurity phase (IP). (b) Magnetic diffuse scattering at 35 K (symbols) after subtraction of the non-magnetic background \cite{background}. The dashed line is a fit of the low-$Q$ side of the scattering to a Lorentzian lineshape, yielding an average in-plane correlation length $L=60(3)$ \AA\ (see text). The solid line is a simulation using a two dimensional scattering model (see text).} 
\end{figure}

Figure \ref{lowQ}(a) shows the high-intensity neutron scattering profile for 0.35 \AA$^{-1} <Q < 1.4$ \AA$^{-1}$ at selected temperatures. At 300 K, only incoherent scattering due to uncorrelated magnetic moments is observed as expected for a paramagnetic phase. At 3 K, Gaussian peaks with widths given by the instrumental resolution are seen at the magnetic Bragg positions consistent with the type-I AFM structure (see above). At intermediate temperatures, both below and above $T_{N1}$, a broad and asymmetric contribution to the scattering is also observed, peaked at the position of the $(001)'$ Bragg reflection. This is more clearly viewed in the data shown in Fig. \ref{lowQ}(b) obtained after a background subtraction procedure \cite{background}. Careful inspection of the data over an extended $Q$-range (not shown) indicates the diffusive scattering at 35 K is almost totally concentrated at low $Q$, which is indicative of the magnetic nature of this signal. On the other hand, isotropic type-I AFM correlations should also lead to a strong contribution at the $(110)'$ position at $Q \sim 1.09$ \AA$^{-1}$. However, attempts to fit the asymmetric profile shown in Fig. \ref{lowQ}(b) with two peaks centered at $(001)'$ and $(110)'$ positions were unsuccessful. In fact, no scattering component centered at the $(110)'$ position is detectable in our experiment. This is consistent with magnetic scattering from 2D AFM square layers (see Fig. \ref{profiles}), in which the $(10)'$ reflection is expected at $Q_0 \sim 0.77$ \AA$^{-1}$ and the $(11)'$ reflection at $Q \sim 1.09$ \AA$^{-1}$ is forbidden. Indeed, the asymmetric shape of the diffusive scattering is characteristic of 2D diffraction \cite{Warren,Zhang}. In the limit of very large 2D layers, the powder-averaged diffraction from truncation rods shows a resolution-limited jump from zero to maximum intensity at $Q=Q_0$ and a slow falloff for $Q>Q_0$ \cite{Warren,Zhang}. For finite in-layer correlation length $L$, scattering also occurs in the lower $Q$ side of the peak, due to particle size broadening. An estimation of the average $L$ is obtained from a fit of the low $Q$ side of the scattering to a Lorentzian lineshape, $I(Q) \propto |f_M(Q)|^2 /[4(Q-Q_0)^2+w_l^2]$ (dashed line in Fig. \ref{lowQ}(b)), where $f_M(Q)$ is the Ru magnetic form factor and $w_l=0.105(5)$ \AA$^{-1}$ at 35 K, yielding $L=2\pi/w_l=60(3)$ \AA. The solid line in Fig. \ref{lowQ}(b) is a simulation for the entire $Q$ interval using a 2D powder scattering model \cite{Warren,Zhang} with the same $L=60$ \AA\ obtained from the Lorentzian fit. Thus, the only free parameter of the simulation is the overall scale factor. While fair agreement of this model with the experimental data is obtained for the slow intensity falloff at the higher $Q$ side, a clear discrepancy at the lowest $Q$'s is noticed and attributed to a presumably broad distribution of 2D cluster sizes in contrast to the monodisperse $L$ assumed in the model \cite{Warren}. 

\begin{figure}
\includegraphics[width=0.7\textwidth]{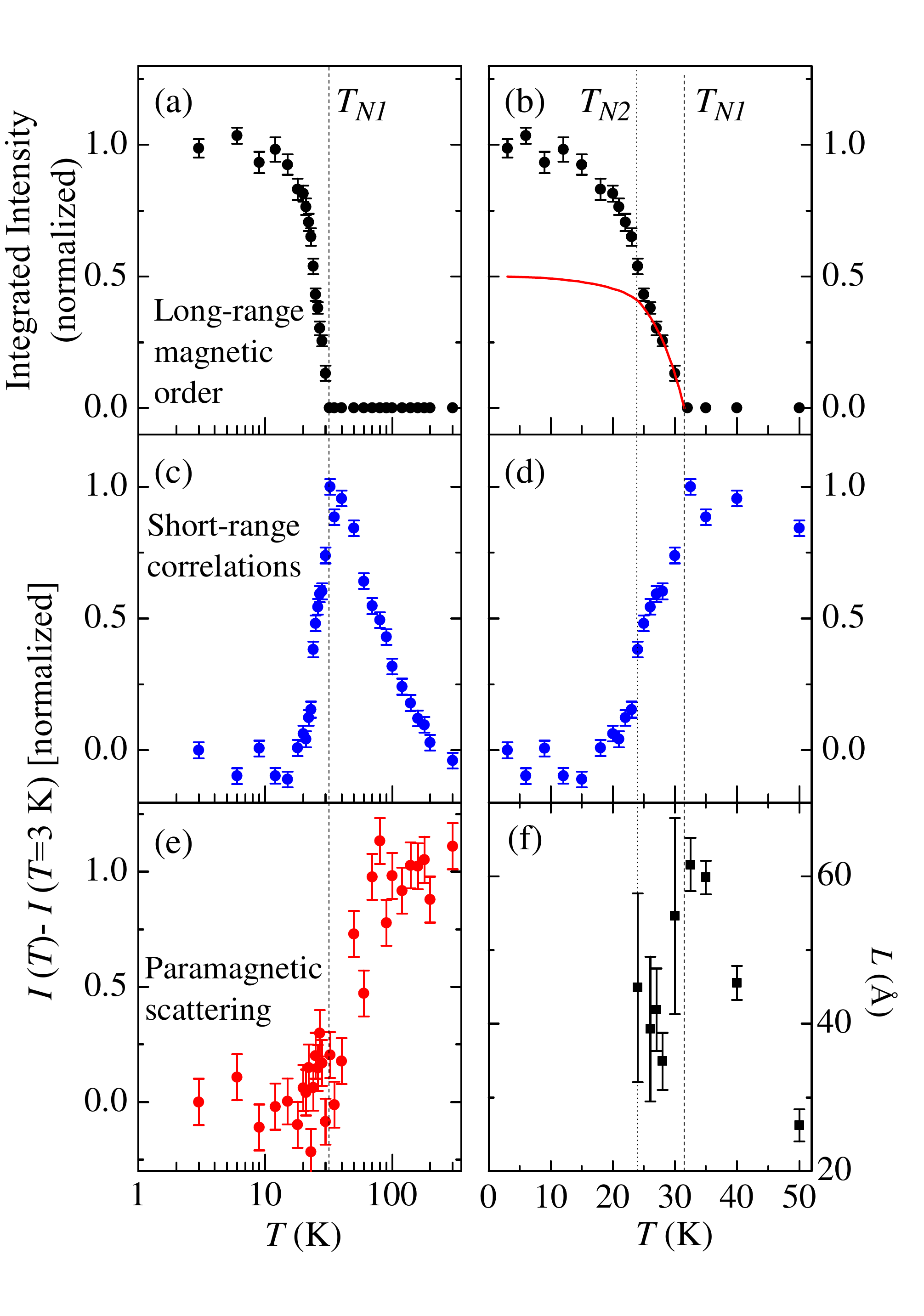}
\caption{\label{Magorder} (a,b) Integrated intensity of the $(110)'$ magnetic reflection on both log and linear temperature scales, highlighting the long-range order parameter. The data in (b) clearly indicate there are two components to the scattering with distinct transition temperatures, and the solid line is a guide to the eye; (c,d) integrated scattering for 0.82 \AA$^{-1}<Q<0.96$  \AA$^{-1}$, capturing signal from short-range magnetic correlations; (e) integrated scattering for 0.35 \AA$^{-1}<Q<0.47$  \AA$^{-1}$, showing the development of weakly correlated spins in the paramagnetic state at high temperatures; (f) in-plane antiferromagnetic correlation length $L$ obtained from half-Lorentzian fits to the difuse scattering component for $Q<0.77$ \AA$^{-1}$ (see text and Figs. \ref{lowQ}(a,b)).} 
\end{figure}

The temperature dependencies of the distinct components of the magnetic scattering are given in Figs. \ref{Magorder}(a-f), taken from the high-intensity data. The intensity of the $(110)'$ magnetic Bragg reflection is given in Figs. \ref{Magorder}(a) and (b). Besides the onset of the long-range magnetic order at $T_{N1}$, a kink in the LRO parameter with an enhancement of the intensity below $T_{N2} = 24$ K is noticed. Figures \ref{Magorder}(c) and (d) show the $T$-dependence of the integrated scattering intensity for 0.82 \AA$^{-1}<Q<0.96$ \AA$^{-1}$, which is an interval dominated by scattering from 2D correlations (see Fig. \ref{lowQ}(b)). These correlations are observed below $\sim 200$ K, and enhance on cooling down to $T_{N1}$. In the ordered phase, the signal from 2D correlations decreases on cooling and nearly disappears below $\sim T_{N2}$. The suppression of the 2D correlations and simultaneous enhancement of the LRO component seem to be related with the phase transition at $\sim T_{N2}$ captured by previous specific heat measurements \cite{Singh}. Figure \ref{Magorder}(e) shows the integrated intensity for 0.35 \AA$^{-1}<Q<0.47$ \AA$^{-1}$, representing the development of incoherent magnetic scattering from uncorrelated paramagnetic moments. The scattering weight from the uncorrelated paramagnetism is progressively suppressed on cooling from 300 K, being transferred to the scattering by 2D correlations. Figure \ref{Magorder}(f) displays $L(T)$, obtained from the Lorentzian fits illustrated in Fig \ref{lowQ}(b), showing a maximum at $T \sim T_{N1}$.

The 2D magnetic correlations in a three-dimensional lattice is a consequence of the inherent geometric frustration associated with the FCC Ru network. In fact, the type-I AFM structure of SYRO may be visualized as a stacking of AFM layers in the $ABAB$ sequence (see Fig. \ref{profiles}). Within a given AFM layer all the nearest-neighbor (n.n.) Ru-Ru interactions are satisfied, while between consecutive layers only half of the n.n. interactions are satisfied and the other half are frustrated. This leads to a cancellation of the coupling energy between consecutive $A$ and $B$ AFM layers that is not restricted to n.n. interactions, remaining valid for quadratic Ru($A$)-Ru($B$) interactions of any range. Therefore, the observed 2D correlations in SYRO are not entirely surprising. In fact, 2D magnetic fluctuations were theoretically predicted for the FCC lattice \cite{Alexander}.

SYRO enters into an interesting state between $T_{N1}$ and $T_{N2}$. In fact, as suggested by the solid line in Fig. \ref{Magorder}(b), the $(110)'$ intensity may be decomposed as the sum of two components with identical weights, one showing a transition at $T_{N1}$ and the other at $T_{N2}$. On the other hand, the intensity ratio between the main magnetic reflections remains constant and the high-resolution diffraction data could be fit well by a simple type-I AFM structure for all temperatures below $T_{N1}$, which at first sight might argue against a LRO magnetic transition at $T_{N2}$. However, these observations can be reconciliated by an intermediate AFM phase for $T_{N2} < T < T_{N1}$, where only alternate AFM square layers are long-range ordered. This phase shows the same magnetic Bragg reflections of the fully ordered type-I AFM phase, however with half of the intensity for each reflection, consistent with Fig. \ref{Magorder}(b). Additional evidence of the partially ordered phase is the persisting signal from short-range correlations between $T_{N1}$ and $T_{N2}$, presumably arising from layers that do not participate in the LRO. A similar partially ordered state has been previously proposed for the FCC antiferromagnet GdInCu$_4$ \cite{Nakamura}, indicating this may be a general trend for this geometry, possibly favored by n.n.n. interactions that may couple alternate but not consecutive  layers (see Fig. \ref{profiles}).

Our results in stoichiometric SYRO call for a rediscussion of some earlier data obtained in Sr$_{2}$YRu$_{1-x}$Cu$_x$O$_6$. Neutron diffraction studies for $x=0.15$ indicated a magnetic signal up to high-$T$, which was attributed to Cu ordering with $T_N$(Cu) $\sim 86$ K in addition to $T_N$(Ru) $\sim 30$ K \cite{Blackstead2}. It is evident, though, that the persisting magnetic scattering above 30 K may be due to correlated Ru moments (see Fig. \ref{lowQ}(a)), and it may not be possible to establish unambiguously whether Cu spins indeed contribute to the ordered magnetism of the doped samples. Also, a relatively small ordered moment of 1.13 $\mu_B$/formula unit was observed in Sr$_{2}$YRu$_{0.85}$Cu$_{0.15}$O$_6$ \cite{Blackstead2}, suggesting that the partially ordered phase revealed here for pure SYRO may be also present in the Cu-doped samples even at low-$T$. Such a phase would allow for alternating long-range magnetically ordered and disordered YRu$_{1-x}$Cu$_x$O$_4$ layers below $T_N$, with no net exchange field from the ordered layers into the disordered ones. Inhomogeneous superconductivity might flourish in the disordered layers, possibly favored by strong two-dimensional magnetic fluctuations.

In summary, neutron powder diffraction experiments in SYRO with quasi-FCC crystal structure reveal 2D AFM correlations above $\sim 24$ K and a partially ordered state between 24 and 32 K. This behavior is argued to be a direct consequence of the inherent geometrical frustration of antiferromagnetism in this lattice, and may host a variety of interesting physical phenomena, including the possibility of alternating layers of magnetic long-range order and superconductivity in Sr$_2$YRu$_{1-x}$Cu$_x$O$_6$. 

This work was supported by FAPESP and CNPq, Brazil, and NSF Grant No. DMR-0805335, USA.

\newpage

\newpage

\begin{table}[t]
\caption{Supplemental Material: Structural parameters of Sr$_2$YRuO$_6$ (space group $P2_1/n$) at selected temperatures obtained from Rietveld refinements with high-resolution data. Relevant bond distances and angles are also given. Statistical uncertainties are shown in parentheses and represent one standard deviation.}

\begin{tabular}{cccc}
\cline{1-4}
\text{$T$} (K) & 3 & 27 & 45 \\
\cline{1-4}
\text{$a$ (\AA)} & 5.75818(8) & 5.75822(12) & 5.75842(12)\\
\text{$b$ (\AA)} & 5.77700(8) & 5.77718(11) & 5.77704(11)\\
\text{$c$ (\AA)} & 8.14743(13) & 8.1473(2) & 8.1475(2)\\
\text{$\beta$ ($^{\circ}$)} & 90.286(1) & 90.287(1) & 90.287(1)\\
\text{$V$ (\AA $^{3}$)}&  271.021(2) & 271.028(3) & 271.038(3)\\
\cline{1-4}
\text{Sr} $(x,y,z)$\\
\text{$x$} & 0.5073(2) & 0.5072(3) & 0.5075(3)\\
\text{$y$} & 0.53050(12) & 0.5306(2) & 0.5304(2) \\
\text{$x$} & 0.2489(2) & 0.2487(3) & 0.2489(2) \\
\text{$U_{iso}$ $\times$ 100 (\AA$^{2}$)}  & 0.25(2) & 0.27(2) & 0.26(2)\\
& & & \\
\text{Y} $(0,1/2,0)$\\
\text{$U_{iso}$ $\times$ 100 (\AA$^{2}$)} & 0.245(13) & 0.22(2) & 0.25(2)\\
& & & \\
\text{Ru} $(1/2,0,0)$\\
\text{$U_{iso}$ $\times$ 100 (\AA$^{2}$)} & 0.245(13) & 0.22(2)& 0.25(2)\\
& & & \\
\text{O(1)} $(x,y,z)$\\
\text{$x$} & 0.2312(2) & 0.2313(3) & 0.2311(3)\\
\text{$y$} & 0.2004(2) & 0.2001(3) & 0.2007(3) \\
\text{$z$} & -0.0351(2) & -0.0348(3) & -0.0347(3) \\
\text{$U_{iso}$ $\times$ 100 (\AA$^{2}$)} & 0.55(4) & 0.50(6) & 0.56(6)\\
& & & \\
\text{O(2)} $(x,y,z)$\\
\text{$x$} & 0.3027(3) & 0.3028(4) & 0.3025(4)\\
\text{$y$} & 0.7295(3) & 0.7298(4) & 0.7297(4)\\
\text{$z$} & -0.0382(2) & -0.0384(2) & -0.0385(2)\\
\text{$U_{iso}$ $\times$ 100 (\AA$^{2}$)} &  0.38(3) & 0.37(5) & 0.34(5)\\
& & & \\
\text{O(3)} $(x,y,z)$\\
\text{$x$} & 0.4311(2) & 0.4312(3) & 0.4306(3)\\
\text{$y$} & -0.0147(2) & -0.0144(3)& -0.0146(3)\\
\text{$z$} & 0.2341(2) & 0.2344(2) & 0.2342(2)\\
\text{$U_{iso}$ $\times$ 100 (\AA$^{2}$)} &  0.40(3) & 0.43(4) & 0.46(4)\\
\cline{1-4}
\text{Ru-O(1) (\AA)} & 1.9529(11) & 1.9511(15) & 1.9538(15)\\
\text{Ru-O(2) (\AA)} & 1.9561(14) & 1.954(2) & 1.956(2)\\
\text{Ru-O(3) (\AA)} & 1.9520(13) & 1.954(2) & 1.953(2) \\
\text{Y-O(1) (\AA)} & 2.2030(11) & 2.204(2)  & 2.201(2) \\
\text{Y-O(2) (\AA)} & 2.2130(14) & 2.215(2) & 2.213(2) \\
\text{Y-O(3) (\AA)} & 2.2019(13) & 2.200(2) & 2.202(2) \\
\text{Ru-O(1)-Y ($^\circ$)} & 157.78(8) & 157.87(11) & 157.94(11) \\
\text{Ru-O(2)-Y ($^\circ$)} & 155.99(7) & 155.96(10) & 155.94(10) \\
\text{Ru-O(3)-Y ($^\circ$)} & 157.40(6) & 157.45(9) & 157.25(9) \\
\cline{1-4}
\text{$R_p$} (\%) & 3.4 & 3.4 & 3.5\\
\text{$R_{wp}$} (\%) & 4.3 & 4.4 & 4.4 \\
\text{$\chi^2$} & 1.92 & 1.90 & 1.90 \\
\cline{1-4}
\end{tabular}
\end{table}

\begin{figure}
\includegraphics[width=0.9\textwidth]{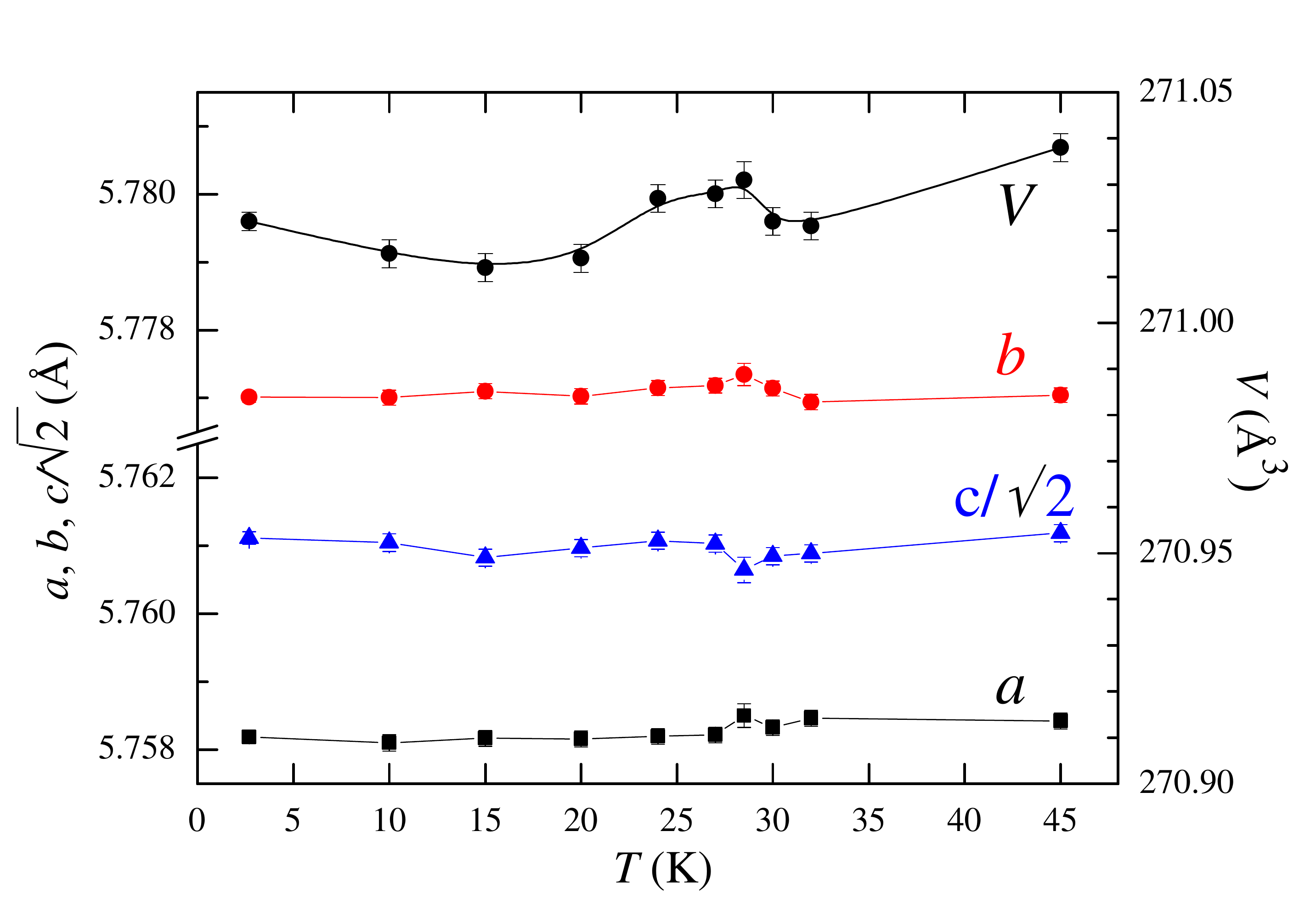}
\caption{\label{Tdep} Supplemental Material: Lattice parameters $a$, $b$, and $c/\sqrt{2}$ and unit cell volume $V$ of Sr$_2$YRuO$_6$ obtained from the refinements using high-resolution neutron powder diffraction data.} 
\end{figure}


\begin{references}
\bibitem{background} The signal at 3 K was used as a reference to subtract the non-magnetic background from the data at higher $T$, since at the lowest $T$ all magnetic scattering is concentrated at the sharp Bragg peaks. This procedure is accurate at the low $Q$'s of Figs. \ref{lowQ}(a,b), where the $T$-dependent phonon contribution to the background can be ignored. A linear extrapolation was employed for the background at the narrow $Q$ intervals around the $(001)'$ and $(100)'$ magnetic Bragg peaks at 3 K.
\end{references}
\end{document}